\documentclass[preprint,12pt]{elsarticle}

\usepackage{amssymb}
\usepackage{amsmath}
\usepackage{parskip}
\usepackage{multirow}

\usepackage{graphicx}
\usepackage{graphics}
\usepackage{float}
\usepackage[caption = false]{subfig}
\usepackage{setspace}

\journal{arXiv}

\begin{document}

\begin{frontmatter}



\title{Confidence Intervals for Finite Difference Solutions}


\author{Majnu John$^{a,}$$^{b}$\footnote{Corresponding author: Department of Mathematics, $308$ Roosevelt Hall, $130$ Hofstra University, Hempstead, NY 11549.  e-mail: {\sf Majnu.John@hofstra.edu}, Phone: +01\,718\,470\,8221, Fax: +01\,718\,343\,1659}, Yihren Wu$^{a}$}

\address{$^{a}$Department of Mathematics, \\Hofstra University, \\Hempstead, NY.}
\address{$^{b}$Center for Psychiatric Neuroscience,\\ The Feinstein Institute of Medical Research,\\Manhasset, NY.}

\begin{abstract}

Although applications of Bayesian analysis for numerical quadrature problems have been considered before, it's only very recently that statisticians have focused on the connections between statistics and numerical analysis of differential equations. In line with this very recent trend, we show how certain commonly used finite difference schemes for numerical solutions of ordinary and partial differential equations can be considered in a regression setting. Focusing on this regression framework, we apply a simple Bayesian strategy to obtain confidence intervals for the finite difference solutions. We apply this framework on several examples to show how the confidence intervals are related to truncation error and illustrate the utility of the confidence intervals for the examples considered.

\end{abstract}

\begin{keyword}


Numerical Analysis, Finite Difference Method, Bayesian Analysis

\end{keyword}

\end{frontmatter}





\noindent

\section{Introduction}

Until the last decade or so, the connections between statistics and numerical analysis had not been explored much. Research work that addressed such connections were few and far in between until recently. Diaconis (1988) was one of the first expository papers that provided examples of connection between Bayesian analysis and numerical methods; to paraphrase Persi Diaconis from that paper, `it may even sound crazy to think that there exists connections between numerical analysis and statistics'. However, such connections \textit{do} exist, as pointed out in Diaconis (1988), and further explored in papers by O'Hagan, Ylvisaker, (O'Hagan (1991), Ylvisaker (1987)) and many others and recently by Conrad and co-authors (Conrad \textit{et al}, 2015) and  Chkrebtii and co-authors (Chkrebtii  \textit{et al}, 2016). For a comprehensive list of papers related to this area, please refer to the references within the last two papers mentioned above. O'Hagan's and Diaconis' work mainly addressed relations of statistics, especially Bayesian statistics, to quadrature problems. Ylvisaker's work was related to optimal statistical design in a general context, but also had specific connections to quadrature problems. All the above work were related to connections between statistical analysis and numerical methods for quadrature or design points (mainly for quadrature); none of them established a connection between statistical analysis and numerical methods for differential equations. Recent papers, by Conrad and co-authors and  Chkrebtii and co-authors were the first two papers that addressed this connection, to the best of our knowledge. The work for the present paper was done independently of the work by Conrad and  Chkrebtii and respective co-authors (-in Summer, 2016 and the authors were not aware of Conrad \textit{et al} and Chkrebtii  \textit{et al} until the writing stage of this paper, in Fall 2016, when they were published). Although our paper is similar in spirit to Conrad \textit{et al} and Chkrebtii  \textit{et al}, the examples that we provide are completely different from those two papers. Our statistical approach is somewhat simpler than theirs; we hope that our work will be a good addendum to the above mentioned papers, and to this new emerging field of research in general.

For many finite difference methods that find an approximate solution $\boldsymbol{\hat{\beta}}$ at discrete points for a differential equation, we may associate a linear regression \begin{equation}\label{eq:1} Y = X\boldsymbol{\beta} + \varepsilon. \end{equation} Here $\boldsymbol{\beta}$ is the unknown (column) vector of original solution values at the respective discrete points, $Y$ is a known (column) vector, $X$ is a known matrix, and $\varepsilon$ is a vector of errors, with $Y$ and $X$ possibly depending on $\boldsymbol{\hat{\beta}}$ and $\varepsilon$ on the truncation error due to the approximation. The type of the linear regression may vary from one finite difference method to the other. Typically using this equation to estimate $\boldsymbol{\beta}$ has limited value in the sense that the estimated $\boldsymbol{\beta}$ will not be not be much different from the approximation, $\boldsymbol{\hat{\beta}}$. However, the regression equation could be utilized to obtain confidence intervals and this has a few important practical merits.

The importance and practical utility of a confidence interval obtained via this approach are mainly due to the fact that the width of the interval depend mostly on the size of the truncation error. One obvious merit is that confidence intervals will give one an idea about the precision of the approximation. In many physical, engineering and finance industry applications, differential equations are solved for daily practical purposes, and hence knowing the confidence interval will guide in making the decisions. Another important practical use is in improving the finite difference approximation itself. If there are segments of the solution where the confidence intervals are much wider compared to other parts of the solution, then increasing the number of design points (that is, decreasing the mesh width) just for that particular segment will lead to a better approximation. In a way, this is an adaptive approach for coming up with a non-uniform grid, with grid points clustered in regions where they are most needed.

The goal of our paper is to explain these ideas more clearly using many different examples of finite difference methods. We start with a basic example. Before we get to the examples, we describe the simple Bayesian statistical approach that we utilize in all these examples.

We use a Bayesian approach to obtain the confidence intervals for the estimate of $\boldsymbol{\beta}$  based on eq.(\ref{eq:1}). We follow the standard approach by first assuming that $\varepsilon$ is distributed $N(0, \sigma^2 I)$ (that is, multivariate normal with mean vector $0$ and variance-covariance matrix $\sigma^2 I$, where $I$ is the identity matrix). We assume a flat prior for $\boldsymbol{\beta}$ and an inverted-Gamma prior, $IG(a, b)$ for $\sigma^2$, so that the posterior distribution for $\boldsymbol{\beta}$ is given by
\[P\left(\boldsymbol{\beta}|Y, X, \sigma^2\right) = N\left((X^{T}X)^{-1}X^{T}Y, \sigma^2(X^{T}X)^{-1}\right). \] In each of the examples below, we draw $50500$ samples from inverse-gamma prior for $\sigma^2$ and then draw the corresponding $\boldsymbol{\beta}$ from the posterior density. The parameters for the inverse-gamma prior were chosen as $a = m/2$ and $b = a + 1$, where $m + 1$ is the number of design points. We throw away the first $500$ $\boldsymbol{\beta}$ draws, and then calculate the $2.5\%$ and $97.5\%$ percentiles from the remaining $50000$ draws to obtain the lower and upper confidence limits for $\boldsymbol{\beta}$. We note here that although we refer to them as confidence intervals, technically the standard terminology is credible intervals. 

\section{A simple example}

We start with a simple finite difference method, slightly adapted from one of the first examples given in LeVeque (2007). We consider the second order ordinary differential equation (ODE)
\[   \begin{array}{l} u''(x) = \sin (x), \;\;\; 0 < x < \pi \\
                      u(0) = 1, \;\;\; u(\pi) = \pi + 1. \end{array} \]
The advantage of working with this simple problem is that we know the exact solution: $\displaystyle u(x) = -\sin (x) + x + 1.$ With a mesh width $\displaystyle h = \pi/(m+1)$ and design points $x_{j} = jh$, we denote by $\hat{U}_{j}$ the approximation to the solution $u(x_{j}), j = 0, 1, \ldots, m+1$, to be obtained via the finite difference scheme. We set $\hat{U}_{0} = 1$ and $\hat{U}_{m+1} = \pi + 1$ to correspond with the boundary conditions. Using a centered difference approximation for $u''(x)$, we may write a finite difference equation \begin{equation} \label{eq:2} \frac{1}{h^2} (\hat{U}_{j-1} - 2\hat{U}_{j} + \hat{U}_{j+1}) = \sin(x_{j}),\;\;\; j = 1, \ldots, m, \end{equation} which leads to a linear system of $m$ equations $\displaystyle A\hat{U} = F$, where \[ A = \left[ \begin{array}{cccccc}
                                                            -2 &  1 &        &   & &               \\
                                                             1 & -2 &  1     &   & &               \\
                                                               &  1 & -2     & 1 & &               \\
                                                               &    & \ddots & \ddots & \ddots &   \\
                                                               &    &        & 1 & -2 & 1          \\
                                                               &    &        &   & 1  & -2 \end{array} \right], \;\;\; F = \left[ \begin{array}{c} \sin(x_{1}) - \frac{1}{h^2} \\
                                                               \sin (x_{2}) \\ \vdots \\ \sin (x_{m-1}) \\ \sin (x_{m}) - \frac{(\pi + 1)}{h^2} \end{array} \right] \;\;\; \mathrm{and}\;\;\;     \hat{U} = \left[ \begin{array}{c} \hat{U}_{1} \\  \hat{U}_{2} \\ \vdots \\ \hat{U}_{m-1} \\ \hat{U}_{m} \end{array} \right].\]
The local truncation error is defined by replacing $\hat{U}_{j}$ with the solution $u (x_{j})$ in the finite difference formula eq.(\ref{eq:2}). So, if $\displaystyle U = [u(x_{1}), \ldots, u(x_{m}) ]^{T}$ denote the actual values of the solution, then $\displaystyle AU - F = \tau $, where $\tau$ is the vector of truncation errors. This equation may be re-written in the form of eq.(\ref{eq:1}) with $Y = F$, $X = A$, $U = \boldsymbol{\beta}$ and $\varepsilon = -\tau$.

With $h = \pi/100$ (that is, $m = 99$), we fit the regression parameter $\boldsymbol{\beta}$, and obtained the confidence intervals using the Bayesian method outlined in the introduction. The results are shown in figure 1. In the first panel (figure 1a), the black curve is the actual solution, $\displaystyle u(x) = -\sin (x) + x + 1$, between $0$ and $\pi$, the blue points are the finite difference approximation $A^{-1}F$ and the red points correspond to the regression parameter fit. For this example, the red points are almost smack on top of the blue points (so that it is hard to distinguish between the two sets of points!) and they are perfectly in alignment with the actual solution curve. The red dashed lines correspond to the confidence interval obtained using the Bayesian fit.

Figures 1b and 1c illustrate the main point that we want to make with this example. In figure 1b, we plot the width of the confidence interval for the design points chosen between $0$ and $\pi$. In the figure we scale it with a factor $1/(2 \times 1.96)$, so that if the errors are normally distributed, then this scaled width correspond approximately to the 'standard error'. The shape of this confidence-interval-width-curve is roughly that of a sine curve. This makes sense because analytically (by applying Taylor series expansion to the $j^{\mathrm{th}}$ row in $\displaystyle AU - F = \tau $) we have the expression for the truncation error as \[ \begin{array}{ccl} \tau_{j} &=& [\displaystyle u''(x_{j}) + \frac{h^{2}}{12}u''''(x_{j}) + O(h^{4})] - \sin (x_{j}) \\ &=& \displaystyle \frac{h^{2}}{12}u''''(x_{j}) + O(h^{4}) \\ &=& \displaystyle -\frac{h^{2}}{12}\sin(x_{j}) + O(h^{4}). \end{array} \]

Since the error term $\epsilon$ in the regression equation used in this example was equal to $-\tau$, the leading term in the error is a scaled sine curve. The leading term in the truncation error, $\displaystyle -(h^{2}/12) \sin (x)$ is plotted in figure 1c. It is easy to see that if we invert the curve in figure 1c, we get the same shape as in figure 1b, although the scale is different. Thus, the main point is rather simple: the width of the confidence interval mainly depends on the truncation error at that point. The points where the confidence interval is relatively wider are the points where the truncation error is relatively larger. This simple point can have important practical utility as illustrated in the next few examples.

\begin{figure}[H]
\begin{center}
\includegraphics[height=3in,width=7in,angle=0]{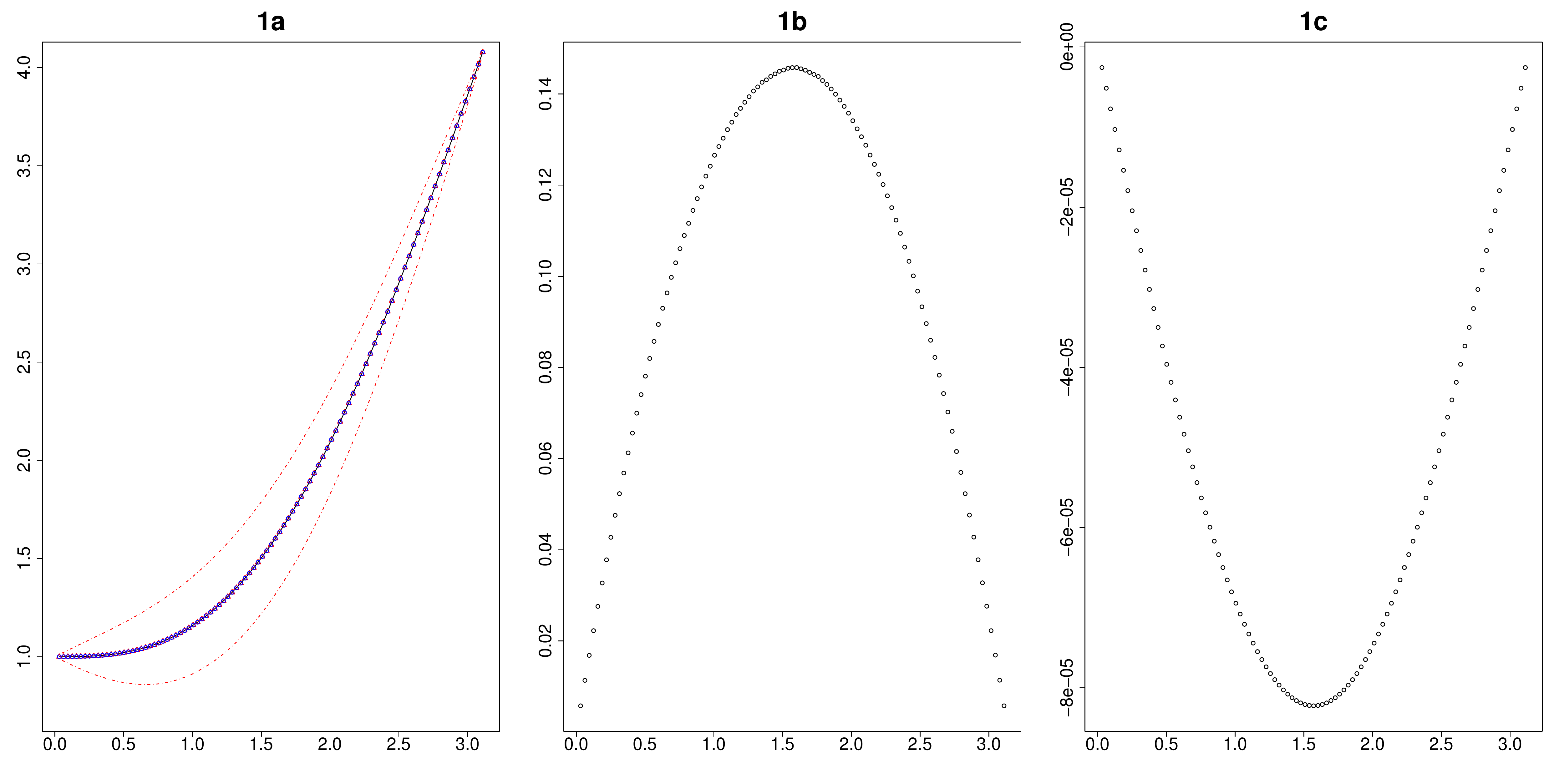}
\caption{$x$-axis in all sub-figures is the interval $[0, \pi]$. 1a) Black curve is the exact solution, blue points are the finite difference solution, red points and the red dashed curves are, respectively, the solution to the corresponding regression equation and the confidence intervals obtained via the Bayesian method. 1b) Width of the confidence interval/($2\times 1.96$), 1c) leading term of the truncation error }
\end{center}
\end{figure}

\section{Nonlinear simple pendulum example}

  Next we consider a nonlinear boundary value problem (BVP) which describes the motion of a simple pendulum with a weight attached to a (massless) bar of length $L$. Ignoring the forces of friction and air resistance, the well-known differential equation for pendulum motion is \[ \theta''(t) = -(g/L)\sin(\theta(t)), \] where $\theta(t)$ is the angle of the pendulum from the vertical at time $t$ and $g$ is the gravitational constant. Taking $g/L = 1$ for simplicity, we have \[ \theta''(t) = -\sin(\theta(t)). \] An initial value problem with the initial position $\theta(0)$ and the initial angular velocity $\theta'(0)$ is most natural for the pendulum problem, and gives a unique solution at all later time points. However, we use the pendulum problem to discuss and illustrate our methods to a BVP. In order to specify the BVP, we set the initial location as $\theta(0) = \alpha$ and the location at a later time point $T$ as $\theta(T) = \beta$. Formally, the 2-point BVP is
  \[   \begin{array}{l}  \theta''(t) = - \sin(\theta(t)) \;\;\;\; \mathrm{for}\; 0 < t < T \\
                         \theta(0) = \alpha, \;\;\;\;\;\;\; \theta(T) = \beta.  \end{array} \]
     For our specific example, we set $\alpha = \beta = 1.2$ and $T = 2\pi$. As pointed out in LeVeque (2007) similar BVPs do arise in more practical situations, for example, trying to shoot a missile in such a way that it hits a desired target.

  Again, following LeVeque (2007), with $\theta_{i} = \theta(t_{i})$, discretization can be done as follows: \[ \frac{1}{h^2}\left(\theta_{i-1} - 2\theta_{i} + \theta_{i+1} \right) + \sin(\theta_{i}) = 0 \] for $i = 1, 2, \ldots, m$, where $h = T/(m+1)$, $\theta_{0} = \alpha$ and $\theta_{m+1} = \beta.$ This can be thought of as a nonlinear system of equations \begin{equation}\label{eq:3} G(\theta_{1}, \ldots, \theta_{m}) = 0, \end{equation} where $G: R^{m} \rightarrow R^{m}$ is the function whose $i^{th}$ component is \[G_{i}(\theta_{1}, \ldots, \theta_{m}) = \frac{1}{h^{2}}\left(\theta_{i-1} - 2\theta_{i} + \theta_{i+1} \right) + \sin(\theta_{i}). \] LeVeque (2007) gives a Newton's iterative scheme for solving eq.(\ref{eq:3}), and for our present example, we assume that this scheme has been used to obtain the solution $\boldsymbol{\hat{\theta}}$ to eq.(\ref{eq:3}). (In this section, we use bold letters to indicate that the solution values or approximations we consider at the design points/grid is a vector.)

  Assume $\boldsymbol{\theta}$ is a solution to the original BVP. Then we may write \[G(\boldsymbol{\theta)} - G(\boldsymbol{\hat{\theta}}) = J(\boldsymbol{\hat{\theta}})(\boldsymbol{\theta} - \boldsymbol{\hat{\theta}}) + O\left( ||\boldsymbol{\theta} - \boldsymbol{\hat{\theta}}||^2\right),  \] where \[J(\boldsymbol{\hat{\theta}}) = \left[J_{ij}(\boldsymbol{\hat{\theta}}) \right]_{m \times m} = \left[ \frac{\partial}{\partial \theta_{j}} G_{i}(\boldsymbol{\hat{\theta}}) \right]; \] that is, \[J_{ij}(\boldsymbol{\hat{\theta}})=\left\{
                              \begin{array}{ll}
                                   \frac{1}{h^2},\;\;\; &\mathrm{if}\;\; j = i - 1\; \mathrm{or} ;\;j = i + 1\\
                                  -\frac{2}{h^2} + \cos(\hat{\theta}_{i}),\;\;\; &\mathrm{if}\;\; j = i \\
                                   0, \;\;\; &\mathrm{otherwise}
                \end{array} \right.\]
    Since $\boldsymbol{\hat{\theta}}$ is a solution to eq.(\ref{eq:3}), we get \begin{equation}\label{eq:4} G(\boldsymbol{\theta)} = J(\boldsymbol{\hat{\theta}})(\boldsymbol{\theta} - \boldsymbol{\hat{\theta}}) + O\left( ||\boldsymbol{\theta} - \boldsymbol{\hat{\theta}}||^2\right).  \end{equation} If we denote the truncation error by $\tau$ and the global error $\boldsymbol{\hat{\theta}} - \boldsymbol{\theta}$ by $E$, then $G(\boldsymbol{\theta}) = \tau$ so that the above equation becomes \[ \tau = J(\boldsymbol{\hat{\theta}})(\boldsymbol{\theta} - \boldsymbol{\hat{\theta}}) + O\left(||E||^2\right). \] Rearranging, \begin{equation}\label{eq:5} J(\boldsymbol{\hat{\theta}})\boldsymbol{\hat{\theta}} = J(\boldsymbol{\hat{\theta}})\boldsymbol{\theta} + \varepsilon, \end{equation} where we have collected all the error terms into the term $\epsilon$. This is a linear regression equation of the form $Y = X\boldsymbol{\beta} + \varepsilon$, where $Y = J(\boldsymbol{\hat{\theta}})\boldsymbol{\hat{\theta}}$, $X = J(\boldsymbol{\hat{\theta}})$ and $\boldsymbol{\beta} = \boldsymbol{\theta}$ (the unknown solution to the BVP at the discrete timepoints). Since the higher order terms for $||E||$ are negligible, the error term $\varepsilon$ is dominated by the truncation error term.

  One more fact to keep in mind regarding pendulum motion is that the corresponding differential equation can be approximated using the linearized equation \[ \theta''(t) = -\theta(t), \] when the amplitude (i.e. $\alpha$) is small. The linearized equation has a harmonic solution $\displaystyle \theta(t) = \alpha \cos (t).$ However, for larger values of the amplitude the harmonic solution is substantially different from the actual solution to the nonlinear differential equation (see, for example, Belendez \textit{et al}, 2007). There is an ``explicit" solution to the IVP in terms of the Jacobi sine function, sn (Belendez \textit{et al}, 2007): \[ \theta(t) = 2\sin^{-1}(k\;\mathrm{sn}(t-t_{0}, k).\] Here $t_{0}$ represent the time that the pendulum reaches the bottom, and \[k^{2} = \frac{1}{2} - \frac{1}{2}\cos (\theta) + \frac{1}{4}\dot{\theta}^2 \] represents the total energy, which is a conserved quantity for this dynamical system. The explicit form allows us to obtain a very accurate solution to our BVP by solving two equations $\theta(0) = 1.2$ and $\theta(2\pi) = 1.2$ for the unknowns $t_{0}$ and $k$. Although this requires a numerical method, the solution will contain an error that is negligible when compared to the error from solving a differential equation, so that for all practical purposes, we may take this numerical solution as the `exact' solution. Our result is $t_{0} = 4.882567374$ and $k = 0.5870761413$; these values are used to plot the blue curve in figure 2.

  We first chose equally spaced design points $t_{i}$ between $0$ and $2\pi$ by setting $h = 4/80$ ($m = 124$). With starting values $\theta_{i}^{[0]} = \alpha \cos(t_{i}) + (\beta - 0.2) \sin (t_{i})$ we ran Newton's iterative scheme for 10 iterations, so that it converged to within a tolerance limit of $10^{-14}$, to obtain the solution for the finite difference scheme (which, keep in mind, is an approximation to the original unknown solution). With this approximation $\boldsymbol{\hat{\theta}}$, we fit the regression model mentioned in the introduction to obtain the confidence intervals.

  \begin{figure}[H]
  \begin{center}
  \includegraphics[height=3in,width=7in,angle=0]{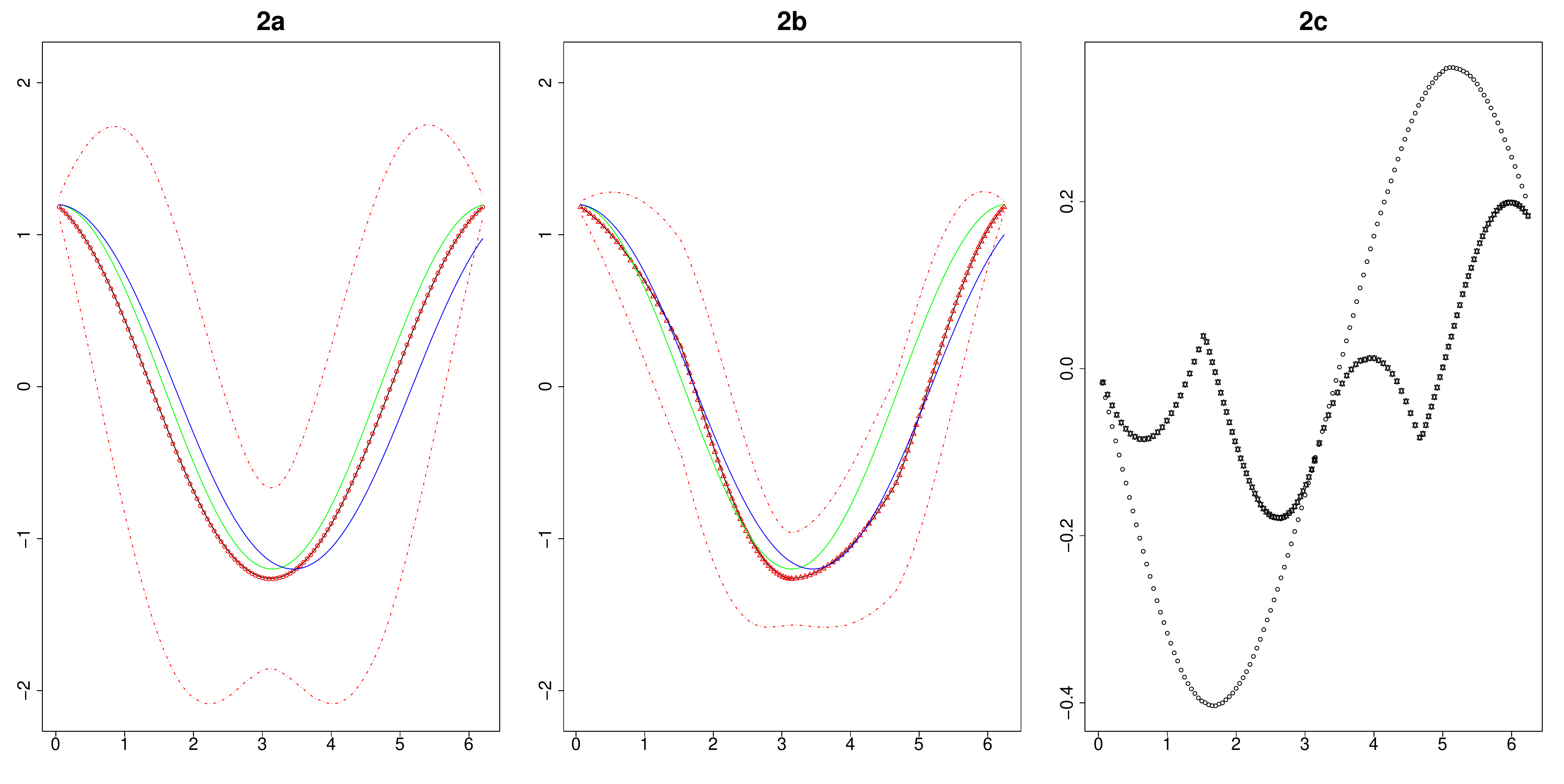}
  \caption{$x$-axis in all sub-figures is the interval $[0, 2\pi]$. 2a) Black curve is the finite difference solution, green curve is the harmonic solution to the linear differential equation, and the blue curve is the exact solution based on Jacobi's elliptic function. The regression fit is plotted as red circles and the corresponding confidence intervals as the red dashed lines. 2b) Colors denote the same thing as for figure 2a. 2c) Circles represent the error corresponding to the finite difference solution in figure 2a, and stars represent the error corresponding to the finite difference solution in figure 2b}
  \end{center}
  \end{figure}

  The results are shown in figure 2a. The finite difference solution obtained using Newton's iterative scheme is plotted as the black solid curve, the harmonic solution to the linear differential equation is plotted as the green curve, and the explicit solution based on Jacobi's sine function is plotted as the blue curve. The regression fit is plotted as red circles and the corresponding confidence intervals as the red dashed lines. The important thing to notice is that the width of the confidence interval varies across the domain, $0 < t < 2\pi$. This suggests that the truncation error is larger in certain sections of the domain, and the accuracy could be improved by increasing the number of design points just in those regions. The confidence interval is relatively wider, for example, in the intervals $[\pi/2, \pi]$ and $[3\pi/2, 2\pi]$. So, we increased the number of design points in those intervals by decreasing $h$ from $4/80$ to $3.3/80$. Since we would like to compare the new results to those presented in figure 2a, we have to ensure that the number of design points is the same in the overall interval $[0, 2\pi]$. Hence we decreased the number of design points in the intervals $[0, \pi/2]$ and $[\pi, 3\pi/2]$ by taking $h = 5.3/80$, so that $m$ remains the same ($m = 124$) as in figure 2a. The revised results are plotted in figure 2b. It is immediately seen that the width of the confidence interval is smaller than that in figure 2a, overall. But, more importantly, the accuracy of the finite difference approximation (the black curve), which overlaps with the Bayesian regression estimate (red triangles), has also increased. This latter fact can be seen more clearly in figure 2c, where the error corresponding to the finite difference solution in figure 2a is plotted as circles and the corresponding error related to figure 2b is plotted as stars. We may observe that the error was reduced not just in the intervals where we increased the number of design points, but overall as well, which is reflective of the fact that the error term in the regression and hence confidence interval in this example depends on \textit{both} the local truncation error $\tau$ and the global error $E$. Thus this example illustrates clearly the helpfulness of calculating and plotting the confidence interval for the finite difference approximation, when we consider adaptively distributing the design points unevenly to reduce error.

  As a last comment related to this example, one may wonder that instead of the eq.(\ref{eq:5}), why not simply consider \[\boldsymbol{\hat{\theta}} = \boldsymbol{\theta} + \varepsilon_{1}, \] for some $\epsilon_{1}$ following a normal distribution (-after all, this is the equation one gets if one divides eq.(\ref{eq:5}) throughout by $\displaystyle J(\boldsymbol{\hat{\theta}})$). The answer is that the error structure based on $\epsilon_{1}$ is not reflective of the truncation error and hence considering this new equation is not very helpful. To illustrate, we fit the confidence interval for the regression estimate with $\displaystyle Y = \boldsymbol{\hat{\theta}}$ and $X = I$, the identity matrix. The results are shown in figure A1 in the appendix. In this case, the confidence intervals are just related to sampling error/variation.


\section{Example with an interior layer for the solution}

  In this example, we consider the nonlinear boundary value problem
  \begin{equation}\label{eq:section.5a}   \begin{array}{l}  \delta u'' + u(u' - 1) = 0 \;\;\;\; \mathrm{for}\; a \leq x \leq b \\
                         \;\;\;\;\;\; u(a) = \gamma_{1}, \;\;\;\;\;\;\; u(b) = \gamma_{2},  \end{array} \end{equation}
  which is a singular perturbation problem since $\delta$ multiplies the higher order derivative. With $\delta = 0$, the reduced first order ODE can enforce only one boundary condition. If $u(a) = \alpha$ is imposed, then the non-trivial solution to the reduced equation is \begin{equation}\label{eq:section.5b} u(x) = x + \gamma_{1} - a \end{equation} and if $u(b) = \gamma_{2}$ is imposed, the corresponding solution is \begin{equation}\label{eq:section.5c} u(x) = x + \gamma_{2} - b. \end{equation} On the other hand, for $ 0 < \delta \ll 1$, the full equation (\ref{eq:section.5a}) has a solution that satisfies both the boundary conditions by following the line given in eq. (\ref{eq:section.5b}) near $x = a$ and the (parallel) line in eq. (\ref{eq:section.5c}) near $x = b$. Connecting these two smooth portions of the solution is a narrow zone (the interior layer) where $u(x)$ is rapidly varying. In this layer, $u''$ is very large and the $\delta u''$ term in eq. (\ref{eq:section.5a}) is not negligible. Singular perturbation analysis can be done to determine the location and the width of the interior layer, and then combining this inner layer with the outer sections (\ref{eq:section.5b}) and (\ref{eq:section.5c}) of the solution one may obtain an approximate solution of the form \[ u(x) \approx \tilde{u}(x) \equiv x - \bar{x} + w_{0} \tanh \left(w_{0}\frac{x-\bar{x}}{2\delta}  \right), \] where \[ w_{0} = \frac{1}{2}(a - b + \gamma_{2} - \gamma_{1}) \;\;\; \mathrm{and} \;\;\; \bar{x} = \frac{1}{2}(a + b - \gamma_{1} - \gamma_{2}).\] (See again Leveque (2007) for details of the derivation of this approximate solution.) The green curve shown in some of the plots below (and appendix) related to this example, correspond to approximate solution based on the above formula. For the specific illustrative examples below we chose $a = 0, b = 1, \alpha = -1$ and $\beta = 1.5$.

  In order to obtain a numerical solution at grid points $\displaystyle x_{i} = a + ih$, $i = 1, \ldots, m$, where $h = (b-a)/(m+1)$, we may discretize eq. (\ref{eq:section.5a}) to obtain the following set of finite difference equations:
  \[ G_{i}(\hat{U}) = \delta \left( \frac{\hat{U}_{i-1} - 2\hat{U}_{i} + \hat{U}_{i+1}}{h^2} \right) + \hat{U}_{i}\left(\frac{\hat{U}_{i+1} - \hat{U}_{i-1}}{2h} - 1 \right) = 0, \] with $\displaystyle \hat{U}_{0} = \gamma_{1}$ and $\hat{U}_{m+1} = \gamma_{2}$. This gives a nonlinear system of equations $G(\hat{U}) = 0$ as in example 2, which can be solved using Newton's method. The values of the perturbation-analysis based approximate solution above at the discrete points $x_{i}$ can be used as starting values for Newton's algorithm. Again, as in example 2, assuming that an approximate solution $\hat{U}$ has been obtained using Newton's iterative scheme, we may set up a linear regression $\displaystyle Y = X\boldsymbol{\beta} + \varepsilon, $ where $Y = J(\hat{U})\hat{U}, \;\; X = J(\hat{U}), \;\; \boldsymbol{\beta} = U$ (the unknown solution to eq. (\ref{eq:section.5a})), and the error term $\varepsilon$ depends mainly on truncation error. Here, $\displaystyle J(\hat{U})$ is an $m \times m$ matrix, with $ij^{th}$ element  \[ J_{ij}(\hat{U})  =  \frac{\partial}{\partial \hat{U}_{j}} G_{i}(\hat{U}) =
  \left\{
                              \begin{array}{ll}
                                   \displaystyle \frac{\delta}{h^2} - \frac{\hat{U}_{i}}{2h},\;\;\; &\mathrm{if}\;\; j = i - 1\; \\
                                  \displaystyle -\frac{2\delta}{h^2} + (\frac{\hat{U}_{i+1} - \hat{U}_{i-1}}{2h} - 1 ),\;\;\; &\mathrm{if}\;\; j = i \\
                                   \displaystyle \frac{\delta}{h^2} + \frac{\hat{U}_{i}}{2h},\;\;\; &\mathrm{if}\;\; j = i + 1\\
                                   \displaystyle \;\;\;0, \;\;\; &\mathrm{otherwise}
                \end{array} \right.\]

  There are two issues that pop up when dealing with this example. The regression approach and the corresponding confidence intervals will be of practical assistance when dealing with both the issues. The first issue is the non-convergence, or convergence to the wrong solution, of Newton's iterative scheme for very small $\delta$. For example, for $\delta = 0.1$, Newton's method converges in about 150 iterations if we use $h = 1/201$ (that is, 200 equally spaced design points). However, when $\delta = 0.01$, for the same equally spaced mesh width, Newton's method doesn't converge even after 500 iterations (see the animation figure uploaded as an ancillary file). In fact, the iterations seem to recycle among the same approximations periodically. Even with a much smaller mesh width $h = 1/1001$ (that is, 1000 equally spaced design points), convergence doesn't occur even in 10000 iterations (plot available on request). Note that Newton's method \textit{does} converge if we choose large number of (e.g. 1000) \textit{unevenly} spaced design points appropriately (-more on this in the next paragraph-), but in this case the convergent solution is far removed from the approximate solution obtained via singular perturbation analysis (see supplementary figure A2). Hence it makes us suspect that Newton's iteration converged to the wrong solution for this non-uniform design.

  The second issue is related to the spacing (uniform vs. non-uniform) of the design points for this example. Using perturbation analysis it could be shown that the width of the interior layer is of order $\delta$ (i.e. very small). Thus, in order to obtain a sufficiently accurate solution, especially with good accuracy for the tiny interior layer, one needs grid points which are sufficiently close enough (hence very large number of grid points). However, if we use a uniformly spaced grid, this would lead to a waste of resources since a large chunk of the solution (that is, the section other than the interior layer) is very smooth and requires only very few grid points to obtain good accuracy. This suggests using a non-uniform grid with design points highly clustered for the interior layer and very sparse for sections other than the interior layer - this could be obtained analytically using singular perturbation analysis. However analytical approaches may not be available all the time. An alternate (non-analytical) approach is based on the confidence intervals, which we illustrate below.

  As we mentioned above, when $\delta = 0.01$ and with 200 equally spaced grid points, Newton's method for the finite difference scheme doesn't converge; so, we stopped at the $500^{th}$ iteration and used the approximation obtained at the last iteration as $\hat{U}$ in the regression equation. The regression fit obtained and the confidence intervals are shown in figure 3a. It is easy to see that the confidence intervals are very wide (and hence the truncation error very large) near the interior layer, compared to other regions. So, we introduced an extra 200 equally spaced points in the subinterval $[0.2, 0.8]$ of the domain with mesh width $(0.8-0.2)/200 = 0.003$. The resulting solution (shown in figure 3b) gets much closer to the approximate solution (green curve) obtained analytically using the perturbation analysis, but still there is a lot of scope for improvement. The width of the confidence interval in figure 3a or figure 3b, if plotted over the domain $[0,1]$, is a bell shaped curve centered at a point near the interior layer. This suggests generating design points from a distribution with a bell-shaped density (e.g. normal distribution). We generated 200 points from a standard normal distribution and rescaled them to fit within the interval $[0,1]$. We used these 200 points in addition to the equally spaced 200 points to run Newton's method till the $500^{th}$ iteration and used the approximation at the last iteration as $\hat{U}$ to fit the regression. The results (shown in figure 3c) are definitely a big improvement over those shown in figures 3a or 3b. We redid the analysis with 400 extra points, and then again with 5000 extra points from the normal distribution, and the results are shown in figures 3d and 3e, respectively. The solution in figure 3e is very close to the approximate solution obtained analytically, and the corresponding confidence intervals are very thin indicating only very small truncation error. In conclusion, this example clearly illustrates how the (width of the) confidence intervals can be used in designing the non-uniform grid for the finite difference method.
  
  We would like to note here that non-convergence that we mentioned above and shown in animation figure A2 in the appendix, is a bit artificial. At each iteration in Newton's method, we have to invert the corresponding matrix $J$. We used the statistical software R for all the numerical computations in this paper. In R, there are 2 ways to invert a matrix: either using the function '\textit{solve}' or using '\textit{chol2inv}'; \textit{solve} is a more generic function which can be used to invert any invertible matrix, while \textit{chol2inv} inverts a symmetric, positive definite square matrix from its Choleski decomposition. In this particular example, both \textit{solve} and \textit{chol2inv} can be used. The non-convergence that we mentioned above is when we use \textit{solve}; when we use \textit{chol2inv} Newton's method converges in about 3 iterations. Nevertheless, since our main goal was to emphasize the utility of confidence intervals, hopefully this example, even though a bit artificial, is illustrative. 

\begin{figure}[H]
\begin{center}
\includegraphics[height=3in,width=7in,angle=0]{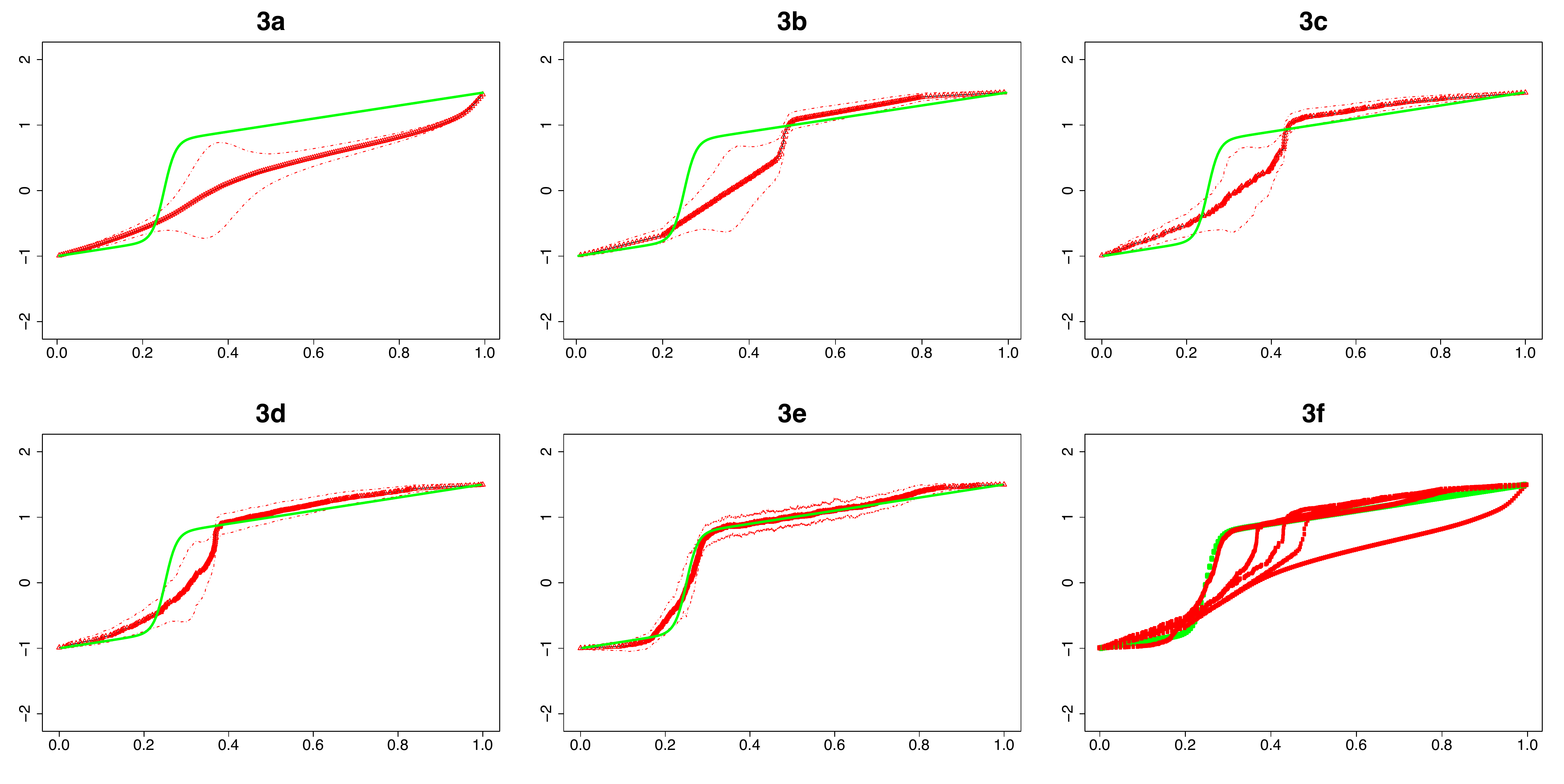}
\caption{$x$-axis in all sub-figures is the interval $[0, 1]$. Green curves in all figures is the approximate solution obtained analytically using singular perturbation method. The solid red curve is the regression fit, and the dashed red lines are the confidence intervals for the regression fit. The number and type (uniform vs. non-uniform) of grid points varied across the subfigures: 2a) 200 evenly spaced grid points with mesh width 0.005; in addition to the grid points in figure 2a, the following sub-figures had extra grid points: 2b) 200 evenly spaced points between 0.2 and 0.8 with width 0.003, 2c) 200 points from a normal distribution scaled to lie in the interval [0, 1], 2d) 400 points from a normal distribution scaled to lie in the interval [0,1] and 2e) 5000 points from a normal distribution scaled to lie in the interval [0,1]. In sub-figure 2f we put together all the regression fits from the previous figures 2a to 2e, going from right to left, so that comparisons could be made. }
\end{center}
\end{figure}

\section{Applications to Black-Scholes equation}

Next we consider discretization of the famous Black Scholes equation for European call option in finance (Hull, 2008). Denoting by $V(S, \tau)$, the value of the option as a function of the underlying asset value (e.g. stock price) and time $\tau$, the corresponding Black Scholes equation is \[ \frac{\partial V}{\partial \tau} + \frac{1}{2} \sigma^2 S^2 \frac{\partial^2 V}{\partial S^2} + r \frac{\partial V}{\partial S} - rV = 0. \] Here $r$, the interest rate, and $\sigma$, the volatility rate are given numbers. Let $T$ denote the expiry time of the option. Applying a change of variables, $t = T - \tau$, so that we have an initial condition, rather than a final condition, we re-write the above equation as \begin{equation}\label{eq:section.bs1} \frac{\partial V}{\partial t} - \frac{1}{2} \sigma^2 S^2 \frac{\partial^2 V}{\partial S^2} - r \frac{\partial V}{\partial S} + rV = 0, \end{equation} and the initial condition as $V(S, 0) = \max (S-E, 0)$, where $E$ is the exercise price. We also have the boundary conditions $V(0, t) = 0$ and $\displaystyle V(S, t) \rightarrow S - E\exp {(-r(T-t))}$ as $S \rightarrow \infty.$ We consider an implicit finite difference method for numerically solving the equation using the following discretization.

We divided the time interval $[0, T]$ into $M$ equally sized subintervals of length $\Delta t$. Technically, $S$ can take values in the infinite interval $[0, \infty)$. In practical situations, for discretization, an upper limit $S_{max}$ is imposed and the finite interval $[0, S_{max}]$ is divided into $N$ subintervals of length $\Delta S$. $S_{max}$ is typically taken as three or four times the exercise price. Based on the above discretization, the rectangle $[0, S_{max}] \times [0, T] = [0, N\Delta S] \times [0, M\Delta t]$ is approximated by a grid $(n\Delta S, m \Delta t), n = 0, \ldots, N, \;\; m = 0, \ldots, M.$ We denote by $v_{n}^{m}$, the value of $V$ at the grid point $(n\Delta S, m \Delta t)$.

In the fully implicit scheme, the derivatives are approximated by \[ \frac{\partial V}{\partial t} (n\Delta S, (m+1)\Delta t) = \frac{v_{n}^{m+1} - v_{n}^{m}}{\Delta t } + O(\Delta t), \] \[ \frac{\partial V}{\partial S} (n\Delta S, (m+1)\Delta t) = \frac{v_{n+1}^{m+1} - v_{n-1}^{m+1}}{2\Delta S } + O((\Delta S)^2), \] \[\frac{\partial^2 V}{\partial S^2} (n\Delta S, (m+1)\Delta t) = \frac{v_{n+1}^{m+1} - 2v_{n}^{m+1} + v_{n-1}^{m+1}}{(\Delta S)^2 } + O((\Delta S)^2). \] Putting it all together, the discretized version of eq. (\ref{eq:section.bs1}) is \[ \left( \frac{v_{n}^{m+1} - v_{n}^{m}}{\Delta t } \right) -  \frac{1}{2} \sigma^2 n^2 (\Delta S)^2 \left( \frac{v_{n+1}^{m+1} - 2v_{n}^{m+1} + v_{n-1}^{m+1}}{(\Delta S)^2 } \right) - rn\Delta S \left( \frac{v_{n+1}^{m+1} - v_{n-1}^{m+1}}{2\Delta S } \right) + rv_{n}^{m+1} = 0,\] \[  n = 1, \ldots, (N-1),\; m = 0, \ldots (M-1).  \] Multiplying throughout by $\Delta t$ and re-arranging terms, the above equation can be written in matrix form as \begin{equation}\label{eq:section.bs2} Av^{m+1} = b^{m},\; m = 0, \ldots, (M-1), \end{equation} where \[ A = \left[ \begin{array}{cccccc}   d_{1} & u_{2} & 0      & \ldots & 0          \\
                        l_{1} & d_{2} & u_{3}  & \ldots & 0          \\
                           0  &\ddots & \ddots & \ddots & \vdots     \\
                      \vdots  &\ddots &        & \ldots & u_{N-1} \\
                           0  &\ldots &        & l_{N-2}& d_{N-1} \end{array} \right],
                        v^{m+1} = \left[ \begin{array}{c} v_{1}^{m+1} \\ v_{2}^{m+1} \\ \vdots \\ v_{N-2}^{m+1} \\ v_{N-1}^{m+1} \end{array} \right],
                            b^{m} = \left[ \begin{array}{c} v_{1}^{m} \\ v_{2}^{m} \\ \vdots \\ v_{N-2}^{m} \\ v_{N-1}^{m} \end{array} \right]
                                  - \left[ \begin{array}{c} l_{0}v_{0}^{m+1} \\ 0 \\ \vdots \\ 0 \\ u_{N}v_{N}^{m+1} \end{array} \right], \]
\[ \begin{array}{rll} d_{n} =& 1 + \beta + \alpha n^2, & n = 1, \ldots, (N-1), \\
                      u_{n} =& -\frac{1}{2}(\beta(n-1) + \alpha(n-1)^2), &n = 2, \ldots, N,  \\
                      l_{n} =&  \frac{1}{2}(\beta(n+1) + \alpha(n+1)^2), &n = 0, \ldots,(N-2), \\\end{array} \] \[ \alpha = \sigma^2 \Delta t, \beta = r \Delta t. \] Here $A$ is an $(N-1) \times (N-1)$ matrix and $v^{m}$, $b^{m}$ are vectors of length $(N-1)$ for all $m$. At each time step, the implicit method provides a linear system of equations
                      based on eq. (\ref{eq:section.bs2}) to be solved. If we denote by $V_{n}^{m}$ the true value of the solution at the grid point $(n\Delta S, m \Delta t)$, $V^{m}$ the corresponding vector, and the truncation error at the $m^{th}$ time step by $e^{m}$, then we have \[ AV^{m+1} = b^{m} + e^{m},\; m = 0, \ldots, (M-1), \] which gives a regression equation at each time step that would help us find a confidence interval for $v^{m}$ (as described in the introduction). We implemented the numerical scheme described above for the following values: $\displaystyle T = 0.25, E = 10, r = 0.1, \sigma = 0.4, S_{max} = 40, N = 200$ and $M = 2000$. Note that the numerical solution that we obtained can be compared with the exact solution for the Black Scholes equation for European call option given by \[ V(S, t) = S\Phi(d_{1}) - Ee^{-r(T-t)}\Phi(d_{2}), \] where the $\Phi(\cdot)$ is the cumulative standard normal distribution function and \[ d_{1} = \frac{\log(S/E) + (r + \frac{1}{2}\sigma^{2})(T-t)}{\sigma\sqrt{T} - t }, \;\;\; d_{2} = \frac{\log(S/E) + (r - \frac{1}{2}\sigma^{2})(T-t)}{\sigma\sqrt{T} - t }. \]

In the figure 4 below, we plot the results for $v^{10}$ (i.e. m = 10). The results and the conclusions for all other time points are similar. In figure 4a, we plot the results based on the exact solution, the numerical solution and the regression estimate, all of which more or less overlap. The confidence intervals for the regression estimate is also plotted. In figure 4b, the absolute error (i.e. $|$exact solution - numerical estimate$|$) is plotted as the black curve. It is easy to note that the absolute error goes up near the exercise price, which is consistent with what is known in the literature. Absolute error may not give the best picture, since the solutions are very near zero to the left of the exercise price, and it is comparatively very high for values far to the right of exercise price. A better error estimate is the relative error, defined as absolute error/exact solution, which is plotted as the blue curve in figure 4b. Relative error is consistently large for prices to the left of the exercise price, and falls off dramatically to the right of the exercise price. We were able to calculate the errors plotted in figure 4b because we know the exact solution. However for many other examples in finance, the exact solution is not available - we have to come up with the error estimates from the numerical solution itself. Our goal is to see whether the confidence intervals proposed in the paper will help us with this query.

The width of the confidence intervals is plotted in figure 4c. As one may see, it doesn't quite match with either of the curves plotted in figure 4b. The width of the confidence intervals steadily goes down to the right as does the relative error, but the shape of the former does not look like that of the latter. In order to obtain a measure that is somewhat similar to the relative error, we may divide the width of the confidence interval by the regression estimate (plotted as the blue curve in figure 4d) or by the sum of the confidence limits (plotted as the black curve in figure 4d). In order to avoid division by zero, we add $1$ to the original value in the denominator. Now, it is clearly seen that the shape of the curves in figure 4d is very similar to that of relative error. Thus, in scenarios like this example, where the value of the solution is very near zero in certain segments and very high for certain others, width of the confidence interval could be used to get an idea of the relative error.

\begin{figure}[H]
\begin{center}
\includegraphics[height=3in,width=7in,angle=0]{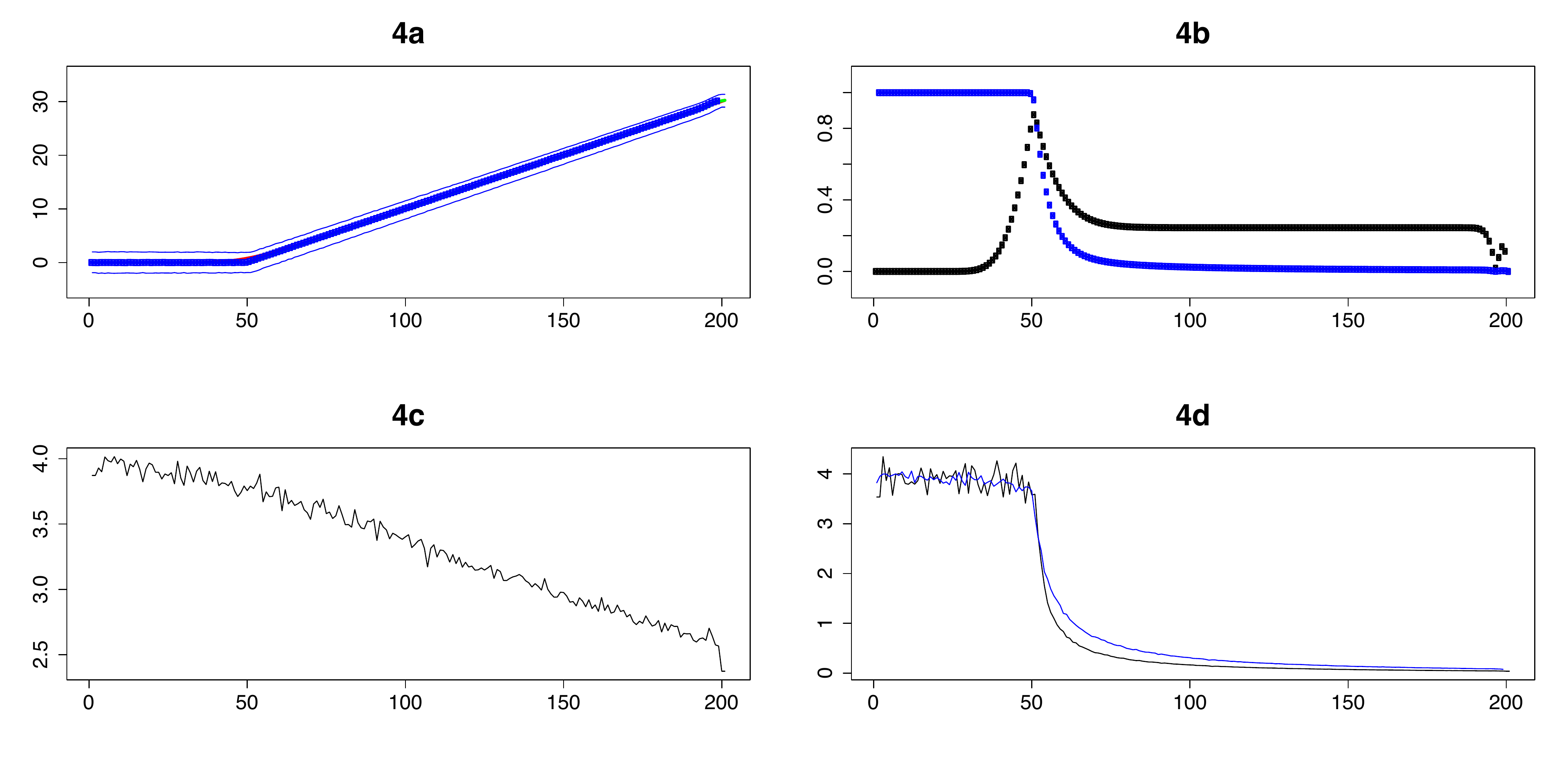}
\caption{In all subfigures, the x-axis represent the grid points $n\Delta S$ for the interval $[0, S_{max}]$; N = 200, n = 0, \ldots, N and $\displaystyle \Delta S = S_{max}/N = 40/200 = 0.2$. The exercise price $E= 10$ corresponds to $n = 50$. Fig 6a: the y-axis represents the option price. Red curve is based on the exact formula; green points based on the numerical solution; blue points based on the regression estimate and the the two blue curves represent the confidence intervals. Fig. 6b: Absolute error is plotted as black points and relative error as blue points. Fig. 6c: Width of the confidence intervals plotted in figure 6a. Fig 6d: Estimates of the relative error based on the width of the confidence intervals, using the formula $(\mathrm{width}\;\;\mathrm{of}\;\;\mathrm{CI})/(D + 1)$, where $D = $ sum of the confidence limits for the black curve and $D =$ regression estimate for the blue curve.}
\end{center}
\end{figure}

\section{Applications in population genetics}

Next we illustrate the usefulness of confidence intervals for an example from population genetics, that uses the Crank-Nicholson implicit finite difference scheme. In population genetics, it is considered that the frequency, $p$, of an allele at a specific genomic location changes over time (that is, from generation to generation). If we assume that the population is sufficiently large and the change in $p$ per generation sufficiently small then the change in $p$ through time may be approximated by a continuous stochastic process. If there are two alleles, $a$ and $A$, at a specific location with frequencies $15\%$ and $85\%$ respectively, in the current generation, then it is possible that at some time in the future (say, $T = 6000^{th}$ generation) the frequency of the allele $A$ is $100\%$. When this happens (that is, when frequency becomes $100\%$) we say that the allele $A$ is fixed. One quantity of interest for population geneticists is the probability of allele fixation at a time/generation $t$, denoted by $u(p_{0}, t)$. Here time (considered synonymous with generation) is a discrete variable $t = 0, 1, \ldots$, where $t = 0$ corresponds to the first generation; $p_{0}$ is the frequency of allele-of-interest in the first generation. Kimura (1964) was one of the first papers that dealt with this probability of fixation; for our illustrative example, we focus on results from Wang and Rannala (2004), where $u(p_{0}, t)$ is given as a solution of the following differential equation (assuming $t$ as continuous for a moment): \[ \frac{\partial u }{\partial t} (p_{0}, t) = \frac{p_{0} (1 - p_{0})}{4 S_{pop}} \frac{\partial^2 u }{\partial p_{0}^2} (p_{0}, t) + s p_{0} (1 - p_{0}) \frac{\partial u }{\partial p_{0}} (p_{0}, t) \] with boundary conditions $u(1, t) = 1\;\; u(0, t) =0.$ Here $S_{pop}$, the size of the population, and $s$, a constant called the 'selection coefficient' are given numbers. We prefer to use the notation $x$ instead of $p_{0}$, so that the above equation in our preferred notation is \begin{equation}\label{eq:section.gen1} \frac{\partial u }{\partial t}  = \frac{x (1 - x)}{4 S_{pop}} \frac{\partial^2 u }{\partial x^2}  + s x (1 - x) \frac{\partial u }{\partial x}.  \end{equation} We also have the initial condition $\displaystyle u(x, 0|x_{0}) = \delta(x - x_{0}), $ where $x_{0}$ is the frequency of the allele in the first generation.

$x$ lies in the interval $[0, 1]$ and we discretize the interval $[0,1]$ with a mesh-width of $\Delta x = 0.0005.$ $t$ ranges from $0$ to $T$ for some large $T$. Since $t$ is interpreted as the $t^{th}$ generation, a natural discretization is based on $\Delta t = 1$. For our example, we follow the above discretization so that the generic grid point is $(n \Delta x, m \Delta t) = (0.0005n, m)$, where $n = 0, \ldots N$, $m = 0, \ldots M$; $N = 2000, M = 6000$, and the value of $u$ at the generic point is represented as $u_{n}^{m}$. For the Crank-Nicholson scheme we have \[ \frac{ \partial u}{\partial t} = \frac{u_{n}^{m+1} - u_{n}^{m}}{\Delta t}, \] \[ \frac{\partial u}{\partial x} = \frac{1}{2} \left(\frac{u_{n+1}^{m} - u_{n-1}^{m}}{2\Delta x} +  \frac{u_{n+1}^{m+1} - u_{n-1}^{m+1}}{2\Delta x}   \right)  = \frac{u_{n+1}^{m+1} + u_{n+1}^{m} - u_{n-1}^{m+1} - u_{n-1}^{m} }{4 \Delta x},\] \[ \frac{\partial^{2} u}{\partial x^{2}} = \frac{1}{2} \left(\frac{u_{n+1}^{m} - 2u_{n}^{m} + u_{n-1}^{m}}{(\Delta x)^2} +  \frac{u_{n+1}^{m+1} - 2u_{n}^{m+1} + u_{n-1}^{m+1}}{(\Delta x)^2}   \right) \] \[ \;\;\;\;\;\;\;\;\;\;\;\;\;\;\;\;\;\;\;\;= \frac{u_{n+1}^{m+1} + u_{n+1}^{m}  - 2u_{n}^{m+1} - 2u_{n}^{m} + u_{n-1}^{m+1} + u_{n-1}^{m}}{2(\Delta x)^2},  \] so that eq. (\ref{eq:section.gen1}) may be approximated by \[ \frac{ \partial u}{\partial t} = \frac{(n\Delta x)(1-n\Delta x)}{4S_{pop}} \left( \frac{u_{n+1}^{m+1} + u_{n+1}^{m}  - 2u_{n}^{m+1} - 2u_{n}^{m} + u_{n-1}^{m+1} + u_{n-1}^{m}}{2(\Delta x)^2} \right) \] \[ \;\;\;\;\;\;\;\;\;\;\;\; + s(n\Delta x)(1-n\Delta x) \left( \frac{u_{n+1}^{m+1} + u_{n+1}^{m} - u_{n-1}^{m+1} - u_{n-1}^{m} }{4 \Delta x} \right). \] Substituting the given values for $S_{pop}, s, \Delta x$ and $\Delta t$ and denoting $\displaystyle \alpha_{n} = 10^{-3}n(1 - 0.0005n)$, we may rearrange the terms in the above approximation to get \[-249.25\alpha_{n} u_{n-1}^{m+1} + (1 + 500\alpha_{n})u_{n}^{m+1} - 250.75\alpha_{n}u_{n+1}^{m+1} \] \[\;\;\;\;\;\;\;\;\;\;\;= 249.25\alpha_{n} u_{n-1}^{m} + (1 -500\alpha_{n})u_{n}^{m} + 250.75\alpha_{n}u_{n+1}^{m},\] \[\;\;\;\;\;\;\;\;\;\;\; n = 1, \ldots, (N-1),\;\; m = 1, \ldots, (M-1),\] which in matrix notation will be \begin{equation}\label{eq:section.gen2} Au^{m+1} = Bu^{m} + b^{m},\;\;\;\mathrm{where} \end{equation} \[ A = \left[ \begin{array}{cccccc}   1 + 500\alpha_{1} & -250.75\alpha_{1} & 0      & \ldots & 0          \\
                        -249.25\alpha_{2} & 1 + 500\alpha_{2} & -250.75\alpha_{2}  & \ldots & 0          \\
                           0  &\ddots & \ddots & \ddots & \vdots     \\
                      \vdots  &\ddots &        & \ldots & -250.75\alpha_{N-2} \\
                           0  &\ldots &        & -249.25\alpha_{N-1}& 1 + 500\alpha_{N-1} \end{array} \right],  \]
 \[ B = \left[ \begin{array}{cccccc}   1 - 500\alpha_{1} & 250.75\alpha_{1} & 0      & \ldots & 0          \\
                        249.25\alpha_{2} & 1 - 500\alpha_{2} & 250.75\alpha_{2}  & \ldots & 0          \\
                           0  &\ddots & \ddots & \ddots & \vdots     \\
                      \vdots  &\ddots &        & \ldots & 250.75\alpha_{N-2} \\
                           0  &\ldots &        & 249.25\alpha_{N-1}& 1 - 500\alpha_{N-1} \end{array} \right],  \] \[ u^{m} = \left[ \begin{array}{c} u_{1}^{m} \\ u_{2}^{m} \\ \vdots \\ u_{N-2}^{m} \\ u_{N-1}^{m} \end{array} \right],
                            b^{m} = \left[ \begin{array}{c} 249.25\alpha_{1}(u_{0}^{m} + u_{0}^{m+1}) \\ 0 \\ \vdots \\ 0 \\ 250.75\alpha_{N-1}(u_{N}^{m} + u_{N}^{m+1}) \end{array} \right]. \] Based on the boundary conditions, we have $u_{0}^{m} = u_{0}^{m+1} = 0$ and $u_{N}^{m} = u_{N}^{m+1} = 1$, so that $b^{m}$ reduces to $\displaystyle \left[  0, 0, \ldots, 0, 501.5\alpha_{N-1}  \right]^{t}.$
If we denote by $U_{n}^{m}$ the actual values of the solution at the grid points and $U^{m}$ the corresponding vector, then from eq. (\ref{eq:section.gen2}) we obtain \[ Bu^{m} + b^{m} = AU^{m+1} + e^{m}, \] where $\displaystyle e^{m}$ denotes the truncation error at the $m^{th}$ time step. With $\displaystyle Y = Bu^{m} + b^{m}$ and $\beta = U^{m+1}$, this is a regression equation of the form mentioned in the introduction, based on which confidence intervals for $U^{m+1}$ can be obtained. In the table below, we present the fixation probability estimates, and the corresponding confidence intervals at the $1000^{th}$ generation when the initial the initial frequency is $0.0005$ and $0.1$, obtained using the above scheme.

\begin{center}
\begin{tabular}{|c|c|c|c|}\hline
\multicolumn{4}{|c|}{Numerical results } \\\hline
  $ p_{0} $  &  fixation probability      &  $95\%$ CI    &  $95\%$ CI     \\
             & ($ u(p_{0}, 1000) $ )      & lower limit   &  upper limit    \\\hline

  0.0005     & 0.000168985    & 0.00  & 1.00       \\\hline
  0.001      & 0.000339403    & 0.00  & 0.6608832  \\\hline
  0.1        & 0.058862310    & 0.00  & 0.3229033  \\\hline

\end{tabular}
\end{center}

\section{Conclusions} Recently, there has been interest in exploring the connections between statistics and numerical analysis related to differential equations. In this paper, we show how some of the common finite difference schemes used to obtain numerical solutions for differential equations can be thought of as a regression problem. Using a simple Bayesian approach to solve the corresponding regression problem, we may obtain confidence intervals for the finite difference solutions. Applying this simple strategy to several basic examples, we show how the confidence intervals are related to truncation error and thus illustrate the utility of the confidence intervals for the examples.

There are limitations to the work presented in this paper, the most obvious being extra (sometimes large amount of) time required to calculate the confidence intervals in addition to the numerical solutions. This would be less of an issue as the speed of computers increase in the future.

\newpage
\appendix
\section{\\Supplementary Figures} \label{App:AppendixA}

\renewcommand{\thefigure}{A\arabic{figure}}
\setcounter{figure}{0}

\begin{figure}[H]
\begin{center}
\includegraphics[height=3in,width=3.35in,angle=0]{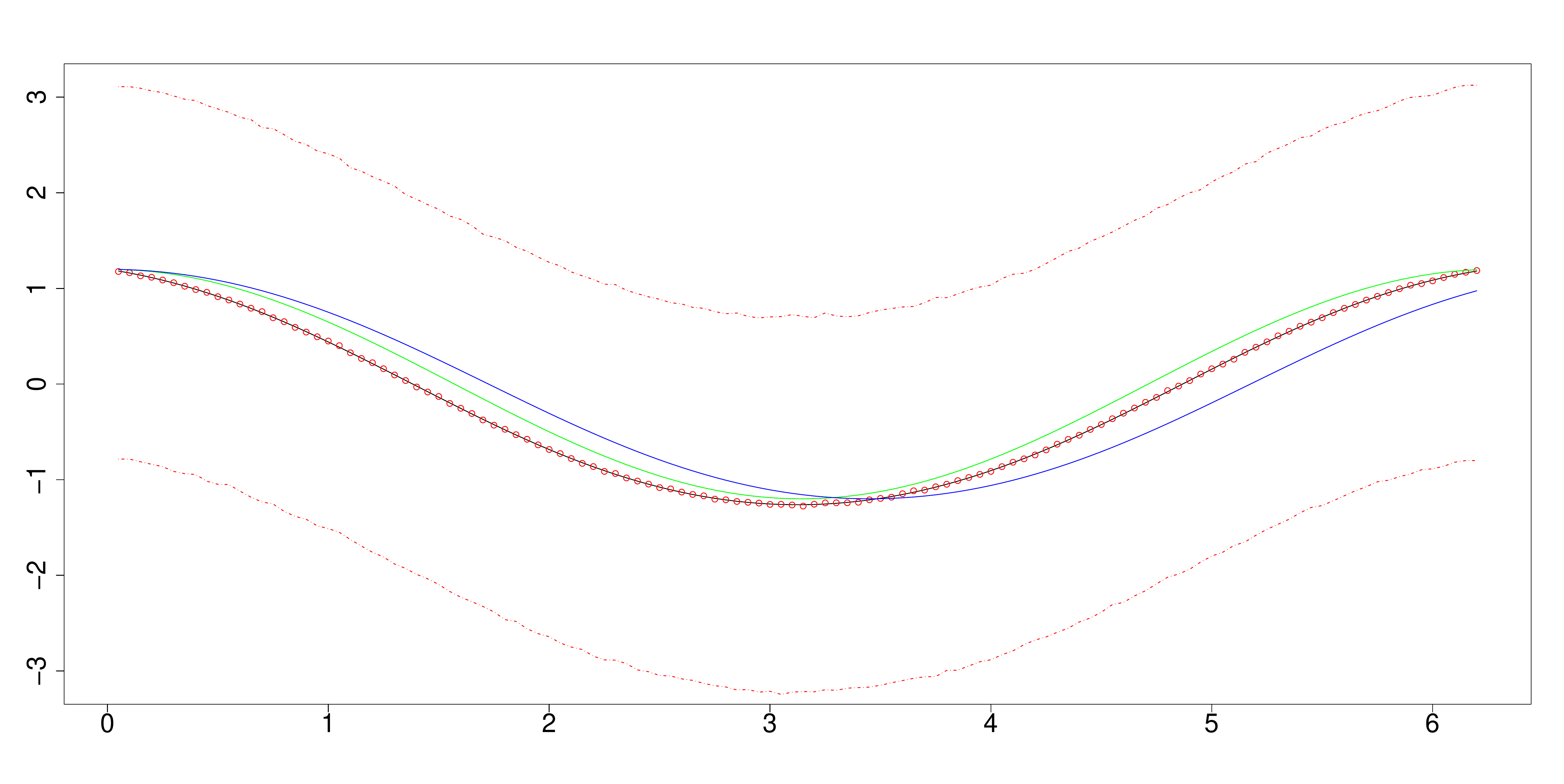}
\caption{$x$-axis is the interval $[0, 2\pi]$. Black solid curve is the finite difference solution obtained using Newton's iterative scheme, green curve is the harmonic solution to the linear differential equation and the exact solution based on Jacobi's elliptic function is plotted as the blue curve. The regression fit is plotted as red circles and the corresponding confidence intervals as the red dashed lines. }
\end{center}
\end{figure}



\begin{figure}[H]
\begin{center}
\includegraphics[height=3in,width=3.35in,angle=0]{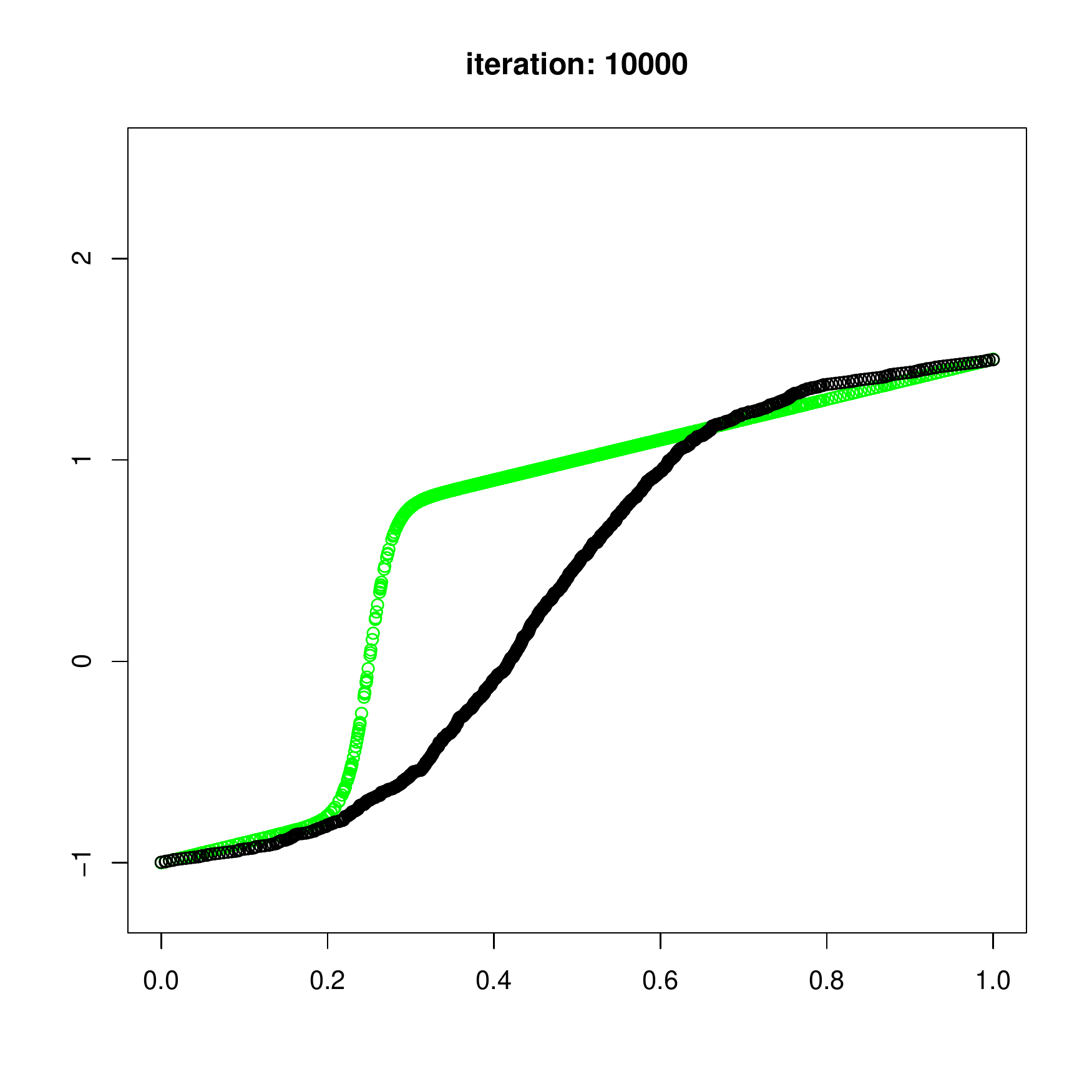}
\caption{Convergence of Newton's iteration, for the difference equation in section 5, to a solution quite different from the approximate solution obtained using singular perturbation analysis, when $\delta = 0.01$ and grid sample size equals 1000 (200 equally spaced and 800 from a normal distribution). Black curve is the finite difference solution at the 10000th iteration of Newton's iterative scheme, green curve is the approximate solution from singular perturbation analysis}
\end{center}
\end{figure}

\end{document}